\documentclass[twocolumn,prl,showpacs,preprintnumbers,amssymb,superscriptaddress]{revtex4}

\usepackage{epsfig, color}
\usepackage{amsthm, amsmath, amsfonts, ae, psfrag}

\newcommand{\upd}{\mathrm{d}}
\newcommand{\bvec}[1]{{\bf\string#1}}

\begin{document}

\title{Density functional theory for Baxter's sticky hard spheres in confinement}

\author{Hendrik Hansen-Goos}
\email{hendrik.hansen-goos@yale.edu}

\affiliation{Yale University, New Haven, Connecticut 06520, USA
}

\author{Mark A. Miller}

\affiliation{University Chemical Laboratory, Lensfield Road, Cambridge CB2 1EW, UK
}

\author{J.S. Wettlaufer}

\affiliation{Yale University, New Haven, Connecticut 06520, USA
}

\affiliation{NORDITA,  Roslagstullsbacken 23, SE-10691 Stockholm, Sweden
}

\date{\today}

\begin{abstract}

It has recently been shown that a free energy for Baxter's sticky hard sphere fluid is uniquely defined within the framework of fundamental measure theory (FMT) for the inhomogeneous hard sphere fluid, provided that it obeys scaled-particle theory and the Percus-Yevick (PY) result for the direct correlation function [Hansen-Goos and Wettlaufer, J. Chem. Phys. {\bf 134}, 014506 (2011)].  Here, combining weighted densities from common versions of FMT with a new vectorial weighted density, we derive a regularization of the divergences of the associated strongly confined limit. Moreover, the simple free energy that emerges is exact in the zero-dimensional limit, leaves the underlying equation of state unaffected, and yields a direct correlation function distinct from the PY expression. Comparison with simulation data for both the bulk pair correlation function and the density profiles in confinement shows that the new theory is significantly more accurate than the PY-based results. Finally, the resulting free energy is applicable to a glass of adhesive hard spheres.

\end{abstract}


\maketitle

Colloidal suspensions are known to exhibit a variety of collective phenomena such as crystallization, gel formation, glass transition and percolation \cite{BuRuPi07,Luetal08,EbWaCa11}. Unlike in molecular systems these properties can be measured not only in terms of bulk sampling such as x-ray diffraction, heat capacity measurements, or mechanical properties, but also on a microscopic level thanks to optical single-particle observation techniques such as confocal microscopy. An ideal theoretical tool to study the structural properties down to the particle size is classical density functional theory (DFT) \cite{Ev79}, an excellent example application being the purely entropic hard-sphere system.  Indeed, within the framework of fundamental measure theory (FMT) \cite{Ro89,KiRo90,KiRo93}, for arbitrary external potentials both the inhomogeneous liquid and the crystalline phases are described by a simple and accurate free energy.  However, a more realistic model of colloidal suspensions must account for attractive interactions.   Such a model is provided by Baxter's sticky hard spheres (SHS) which interact via an infinitely narrow and deep attractive well, the limit being taken such that the second virial coefficient remains finite \cite{Ba68}. The model is particularly appealing because it is soluble analytically within the Percus Yevick (PY) approximation. The PY results obtained by Baxter provide a framework to interpret experimental phase diagrams of colloidal suspensions, including the liquid-liquid transition, crystallization, percolation and, to some extent, the glass transition, revealing the existence of attractive and repulsive glasses \cite{Phetal02}. Quantitative interpretation of single-particle observations in colloidal suspensions rely on a theoretical approach capable of resolving the microscopic structure of Baxter's SHS as accurately as does FMT for hard spheres. Previous DFT approaches relied on Taylor expansions of the free energy around the homogeneous bulk fluid which use the PY result for the direct correlation function while crudely approximating higher order terms. Recently, using a set of weighted densities from FMT for the hard-sphere fluid \cite{HaWe11},  a first non-perturbative DFT was derived. When applied to density profiles near a planar wall, or the SHS confined between two parallel planar hard walls, the theory was shown to be a significant improvement over previous approaches. However, based on experience with Rosenfeld's original FMT for the hard-sphere fluid, it is expected that the FMT for the SHS would exhibit an unphysical divergence for strongly confined fluids, the most extreme test being the zero-dimensional (0D) limit.

In this Letter, we recast the original FMT for the SHS, which is based on Kierlik's and Rosinberg's scalar weighted densities \cite{KiRo90,KiRo93}, by combining Rosenfeld's vectorial weighted density \cite{Ro89} with a new vectorial weighted density.  In consequence, we identify the origin of the divergence and regularize it in the spirit of Schmidt {\em et al.}'s approach to Rosenfeld's FMT for the hard-sphere system \cite{Roetal97}. Rather than deriving a cumbersome expression by conserving the underlying PY correlation function, we seek a {\em simple} density functional.  When the stickiness of the particles is strong, the modified theory (exact in the 0D limit) provides a better description of the pair-correlation function than does the PY-based FMT, while the accurate results at moderate stickiness are left virtually unchanged. For the SHS confined in a spherical cavity, the new theory is significantly better than its predecessor. Our free energy can be used to study a wide range of the structural properties of inhomogeneous colloidal suspensions with short-ranged attractions. In particular, this is the first free energy applicable to the glassy state where confinement due to caging is ubiquitous. In fact, it may overcome previous theoretical pathologies associated with Baxter's adhesive hard-sphere treatment of the glass transition.  These problems were based on using the PY direct correlation function alone \cite{Foetal00}, rather than the complete free energy provided here.

Consider a fluid of monodisperse hard spheres with radius $R$.  We define the locally averaged packing fraction $n_3(\bvec{r}) = \int\upd \bvec{r}'\rho(\bvec{r}')\Theta(R-|\bvec{r}-\bvec{r}'|)$ with weighted densities $n_2(\bvec{r}) = \frac{\partial}{\partial R} n_3(\bvec{r})$ and $\bvec{n}_2(\bvec{r}) = -\nabla n_3(\bvec{r})$. Rosenfeld's excess free energy density (measured in units of $k_B T$) is then  $\Phi_{\text{HS}} = \Phi_1 + \Phi_2 + \Phi_3$ with $\Phi_1 =  -  n_2 \ln(1-n_3)/(4\pi R^2)$, $\Phi_2 = (n_2^2 - \bvec{n}_2\cdot\bvec{n}_2)/[4\pi R (1-n_3)]$, and $\Phi_3 = (n_2^3-3 n_2 \bvec{n}_2\cdot\bvec{n}_2)/[24\pi(1-n_3)^2]$ \cite{Ro89}. While $\Phi_{\text{HS}}$ has been extremely successful in describing an {\em unconfined} fluid, with an inhomogeneity caused, say, by a planar wall, it is problematic when the fluid is highly confined (see \cite{Ta00} and refs.~therein). For the most extreme confinement, the density distribution becomes simply a delta-peak, i.e., $\rho^{\text{0D}}(r) = \eta \delta(r)$, where $r = |\bvec{r}|$, and $\eta$ is the packing fraction ($0\le \eta \le 1$). The weighted densities for this distribution are $n_3^{\text{0D}}(r) = \eta\Theta(R-r)$, $n_2^{\text{0D}}(r) = \eta\delta(R-r)$, and $\bvec{n}_2^{\text{0D}}(\bvec{r}) = \eta\delta(R-r) \hat{\bvec{r}}$, where $\hat{\bvec{r}}=\bvec{r}/r$. The integration of $\Phi_1$ can be readily performed yielding $\int\upd\bvec{r} \Phi_1 = \eta + (1-\eta) \ln(1-\eta)$ which is the exact result for a 0D cavity holding an average number of particles $\eta$  \cite{Roetal97}. Examining $\Phi_{\text{HS}}$ further for such a cavity shows that $\Phi_2$ vanishes (as it should) while $\Phi_3$ exhibits a strong negative divergence. This is due to the factor $3$ in front of the vector term in $\Phi_3$ which would have to be unity for its scalar and vectorial terms to cancel. However, the resulting expression for the excess free energy does not yield the PY direct correlation function, thereby destroying 
 the excellent experimental agreement of Rosenfeld's $\Phi_{\text{HS}}$ for the HS fluid. Several routes have been devised to regularize $\Phi_3$. A particularly appealing approach consists of introducing a new weighted density $\bar{n}_2 = n_2-(\bvec{n}_2\cdot\bvec{n}_2)/n_2$ that reduces to $n_2$ in the bulk fluid (where the vectorial weighted densities vanish as $n_3$ is spatially uniform), and vanishes in the 0D limit (where $\bvec{n}_2\cdot\bvec{n}_2= n_2^2$). Using $\bar{n}_2$, a regularized excess free energy density proposed by Schmidt {\em et al.} \cite{Roetal97} is
\begin{equation}
  \bar{\Phi}_{\text{HS}} = - \frac{n_2 \ln(1-n_3)}{4\pi R^2} + \frac{n_2 \bar{n}_2}{4\pi R(1-n_3)} + \frac{\bar{n}_2^3}{24\pi(1-n_3)^2} \, .
\end{equation}

From the properties of $\bar{n}_2$ it is immediately clear that $\bar{\Phi}_{\text{HS}}$ is exact for a 0D cavity while the free energy of the bulk fluid is unaffected. However, the third term differs by order $[(\bvec{n}_2\cdot \bvec{n}_2)/n_2^2]^2$ from that of $\Phi_{\text{HS}}$, which is very small for moderately inhomogeneous fluids. In fact, it affects only the direct correlation functions of order 4 and higher whereas the PY direct correlation function (which is second order) is preserved.  Both unconfined and confined fluids are extremely well described by $\bar{\Phi}_{\text{HS}}$; accurately predicting solid-liquid coexistence  in hard sphere systems whereas $\Phi_{\text{HS}}$ dramatically overstabilizes the crystal due to the negative divergence for strongly peaked density distributions. However, Tarazona's modification of $\Phi_3$, which uses a tensorial weighted density \cite{Ta00},  performs better with regard to the width of the density peaks in the crystal. 

How does the picture change as we introduce Baxter's attractive surface interaction into the system?
 Rosenfeld's set of weighted densities (i.e.~$n_3$, $n_2$, and $\bvec{n}_2$) is known to be insufficient for constructing any reasonable free energy density for the SHS fluid because they cannot yield the $\delta$-function within the direct correlation function  \cite{HaWe11}. However, when $n_3$ and $n_2$ are supplemented by the weighted density $n_1(\bvec{r}) = \frac{1}{8\pi}\frac{\partial}{\partial R} n_2(\bvec{r})$, the set is sufficient to construct a free energy density $\Phi_{\text{SHS}} = \Phi_{\text{HS}} + \Phi_{\text{S}}$ for Baxter's SHS fluid where $\Phi_{\text{S}}=n_1\phi_1/R + n_2\phi_2/(2\pi R^2)$ is uniquely defined by (a) requiring the density functional to yield the PY direct correlation function and (b) imposing consistency with scaled particle theory \cite{HaWe11}. The dimensionless coefficients $\phi_1$ and $\phi_2$ are functions of $x = \frac{R n_2}{1-n_3}$, which in the bulk fluid is related to the packing fraction $\eta$ via $x = \frac{3\eta}{1-\eta}$.  They are obtained by solving $\phi_1'(x) = - 2 \tilde{y}_\sigma(x)/x$ and $[x^2 \phi_2'(x)]' = \tilde{y}_\sigma(x)^2/2 - x \tilde{y}_\sigma(x) + x \tilde{y}'_\sigma(x)$, where the integration constants are chosen such that $\phi_1$ and $\phi_2$ vanish for $x\to 0$. Here $\tilde{y}_{\sigma}=\eta y_{\sigma}/\tau$, where $y_{\sigma}$ is the cavity function at contact, and is obtained as the smaller of the two solutions to $\tilde{y}_{\sigma}^2-(4x+12\tau)\tilde{y}_{\sigma}+2x(2+x)=0$.

In order to regularize the divergences in the 0D limit we rewrite $\Phi_{\text{S}}$ using vectorial weighted densities so that the resulting functional  $\bar{\Phi}_{\text{S}}$ vanishes. This guarantees that $\bar{\Phi}_{\text{HS}}+\bar{\Phi}_{\text{S}}$ is exact for a cavity that can hold only one particle at a time, and 
thus the corresponding free energy is unaffected by the sticky interaction. We use the identity $n_1(\bvec{r}) = n_2(\bvec{r})/(4\pi R) + (\nabla\cdot\nabla)n_3(\bvec{r})/(8\pi)$ \cite{KiRo93} in $\Phi_{\text{S}}$, and for reasons that will become clear below, we split the term $n_1\phi_1/R$ in half and substitute for $n_1$ in only one of the terms. Moreover, because $\Phi_{\text{S}}$ is always integrated on $\mathbb{R}^3$, we can apply Green's first identity and neglect the boundary integral. The result is $\Phi_{\text{S}} = \Phi_{\text{S1}}+\Phi_{\text{S2}}$, where $\Phi_{\text{S1}}= \left(\phi_{12} n_1 n_2-  \phi_{12}^{\text{v}}\bvec{n}_1\cdot \bvec{n}_2\right)/(1-n_3)$ and $\Phi_{\text{S2}}= \left[\phi_{222} n_2^3-2 \phi_{222}^{\text{v}} n_2 (\bvec{n}_2\cdot \bvec{n}_2)\right]/[4\pi(1-n_3)^2]$. The coefficients $\phi_{12} = \phi_1/(2x)$, $\phi_{12}^{\text{v}} = \phi_1'/2$, $\phi_{222} = (\phi_1/2 + 2 \phi_2)/x^2$, and $\phi_{222}^{\text{v}} = \phi_1'/8$ are functions of $x = \frac{R n_2}{1-n_3}$, and the new vectorial weighted density is $\bvec{n}_1(\bvec{r}) = -\nabla n_2(\bvec{r})/(8\pi)$.

\begin{figure}[tbp]
    
  \includegraphics[height = 4.5cm]{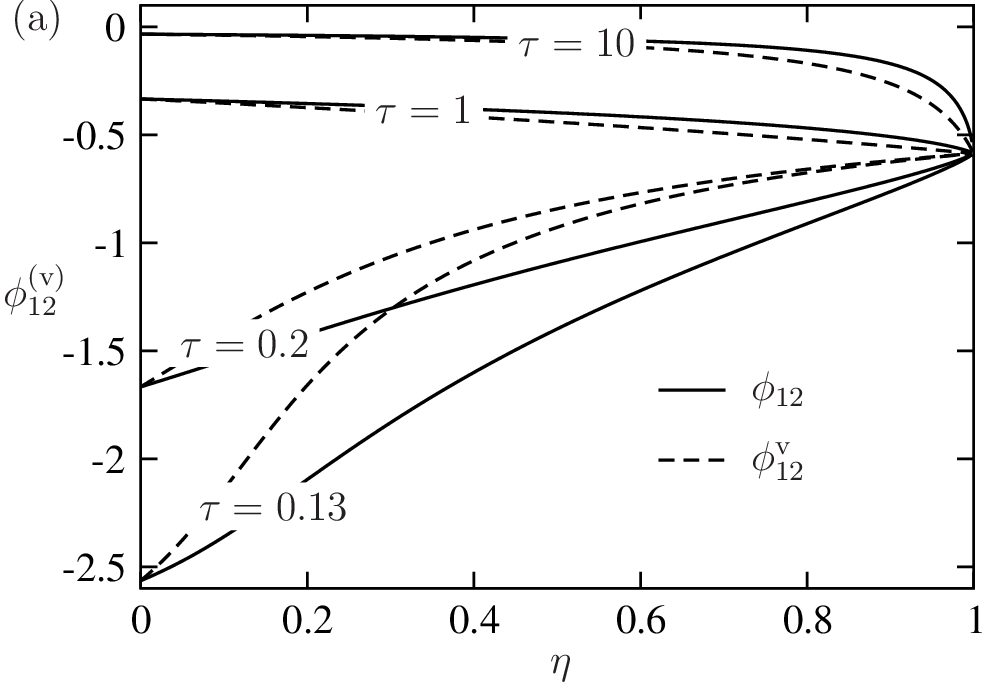} 

  \vspace{0.5cm}
  
  \includegraphics[height = 4.5cm]{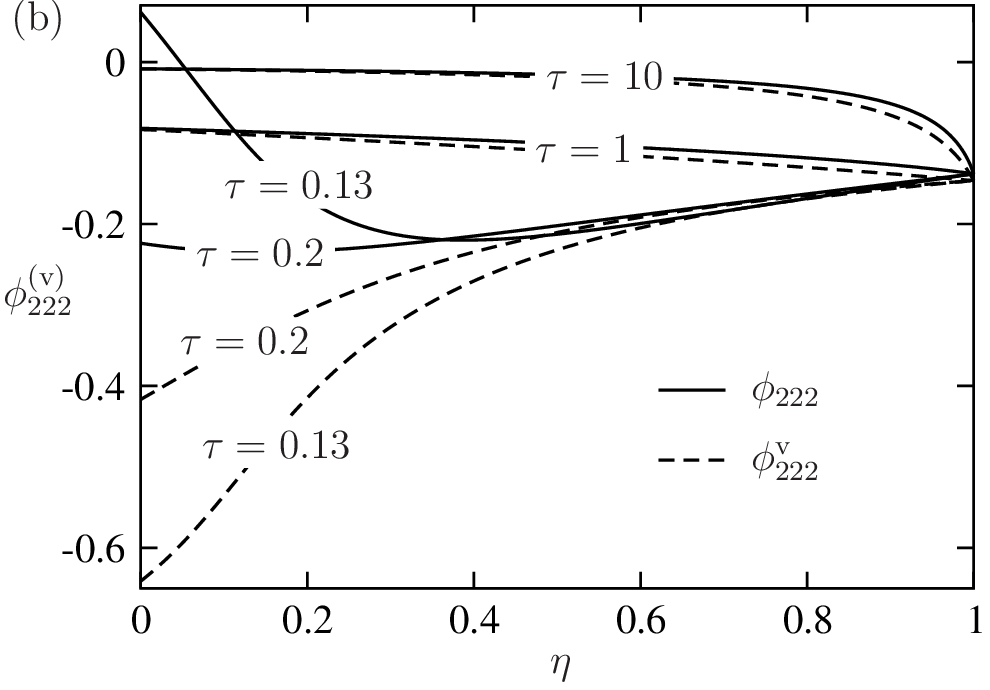} 

  \caption{Coefficient functions in the attractive part $\Phi_{\text{S}}$ of the excess free energy density for SHS with different Baxter parameters $\tau$. In $\Phi_{\text{S}}$ the coefficients are functions of $x=\frac{R n_2}{1-n_3}$, which in the homogeneous bulk fluid is related to the packing fraction $\eta$ via $x=\frac{3\eta}{1-\eta}$. The non-divergent expression $\bar{\Phi}_{\text{S}}$, Eq.~\eqref{eq_PhiSreg}, relies on replacing $\phi_{12}^{\text{v}}$ by $\phi_{12}$ and $\phi_{222}^{\text{v}}$ by $\phi_{222}$ in $\Phi_{\text{S}}$.}\label{fig_phis}
  
\end{figure}

Replacing $n_1$ only in half of the split term $n_1\phi_1/R$ has two important consequences. Firstly, it insures that the function $\phi_{222}$ remains finite in the limit $x\to 0$.  Secondly, it leads to a symmetry between $\phi_{12}$ and $\phi_{12}^{\text{v}}$ that is reflected in their equality in the limits $x \to 0$ and $x\to\infty$, where $\phi_{12}(0)=\phi_{12}^{\text{v}}(0)=-\frac{1}{3\tau}$ and  $\phi_{12}(\infty)=\phi_{12}^{\text{v}}(\infty)=\sqrt{2}-2$. However, comparing $\phi_{12}$ with $\phi_{12}^{\text{v}}$ for any $x$ reveals their differences (Fig.~\ref{fig_phis}). When we replace $\phi_{12}^{\text{v}}$ with $\phi_{12}$, $\Phi_{\text{S1}}$ is unaffected in the bulk fluid. Above we calculated the weighted densities $n_3^{\text{0D}}$, $n_2^{\text{0D}}$ and $\bvec{n}_2^{\text{0D}}$ in the 0D limit.  These are now supplemented by $n_1^{\text{0D}}(r) = \eta\delta'(R-r)/(8\pi)$, and $\bvec{n}_1^{\text{0D}}(\bvec{r}) = \eta\delta'(R-r) \hat{\bvec{r}}/(8\pi)$, from which we have $n_1^{\text{0D}}n_2^{\text{0D}}-\bvec{n}_1^{\text{0D}}\cdot \bvec{n}_2^{\text{0D}}\equiv 0$. Hence, this modification regularizes $\Phi_{\text{S1}}$. Note, that because $\phi_{12}^{\text{v}}-\phi_{12}$ vanishes as $x^{-1}\ln x$ for $x\to\infty$ then even if $\phi_{12}^{\text{v}}$ and $\phi_{12}$ take the same values in this limit it is insufficient to insure that $\Phi_{\text{S1}}$ vanishes in the 0D limit, where large values of $x$ occur due to the delta-function in  $n_2^{\text{0D}}$. Furthermore, larger powers of the vectorial weighted densities could be used to modify the terms $\phi_{12} n_1 n_2/(1-n_3)$ and $\phi_{12}^{\text{v}}\bvec{n}_1\cdot \bvec{n}_2/(1-n_3)$ individually while maintaining for example the PY direct correlation function which underlies $\Phi_{\text{S}}$. However, the symmetry of $\phi_{12}$ and $\phi_{12}^{\text{v}}$ motivates the simple remedy introduced here. The value of giving up Baxter's PY result for the direct correlation function is assessed in terms of the performance of the modified functional in comparison to simulations. 

The regularization of $\Phi_{\text{S2}}$ requires two steps. We replace $\phi_{222}^{\text{v}}$ by $\phi_{222}$ and note that even for asymptotic values of the argument $x$, $\phi_{222}(0) = -\frac{1}{12\tau}+\frac{1}{648\tau^3}$ while $\phi_{222}^{\text{v}}(0)=-\frac{1}{12\tau}$; and $\phi_{222}(\infty) = -\frac{1}{3}(\sqrt{2}-1)\simeq -0.138$ while $\phi_{222}^{\text{v}}(\infty)=-\frac{1}{4}(2-\sqrt{2})=-0.146$, these function differ (see Fig.~\ref{fig_phis}).  Therefore, our substitution is justified by the fact that it becomes exact in the limit of low density ($x\to0$) {\em and} weak ``stickiness'' ($\tau\to\infty$). Like $\Phi_{\text{S1}}$, $\Phi_{\text{S2}}$ in the bulk fluid remains unaffected by the substitution.

After replacing $\phi_{222}^{\text{v}}$ by $\phi_{222}$,  $\Phi_{\text{S2}}$ is still divergent. This is a consequence of the factor $2$ in front of the vectorial term, which comes from matching the $x\to 0$ and $\tau\to\infty$ limits of $\phi_{222}$ and $\phi_{222}^{\text{v}}$. In analogy to $\bar{\Phi}_{\text{HS}}$ where the term $n_2^3-3n_2(\bvec{n}_2\cdot\bvec{n}_2)$ is regularized by replacing it with $\bar{n}_2^3$ we now regularize $n_2^3-2n_2(\bvec{n}_2\cdot\bvec{n}_2)$ by substituting $n_2\bar{n}_2^2$. Introducing the modified weighted density $\bar{n}_1 = n_1- (\bvec{n}_1\cdot\bvec{n}_2)/n_2$ (which equals $n_1$ in the bulk fluid and vanishes in the 0D limit) we can write down the regularized version of $\Phi_{\text{S}}$ as
\begin{equation}
\label{eq_PhiSreg}
 \bar{\Phi}_{\text{S}} =\frac{\phi_{12}\bar{n}_1 n_2}{1-n_3}
   + \frac{\phi_{222}n_2 \bar{n}_2^2}{4\pi(1-n_3)^2} \, ,
\end{equation}
where $\phi_{12}$ and $\phi_{222}$ are functions of $x = \frac{R n_2}{1-n_3}$.

\begin{figure}[tbp]
    
  \includegraphics[height = 4.5cm]{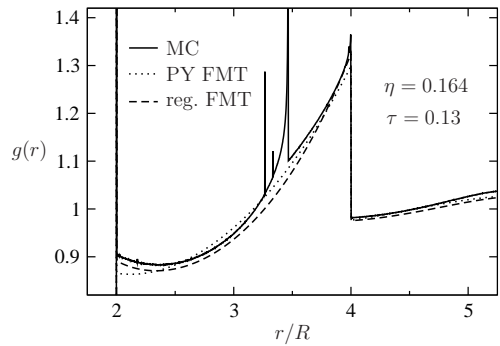} 

  \caption{Pair-correlation function $g(r)$ of a SHS fluid with $\tau=0.13$ and $\eta=0.164$ from MC simulations (solid line) compared to results from the PY-based FMT (dotted line) and the regularized FMT (dashed line).}\label{fig_gr}
  
\end{figure}

As a first test of this regularized free energy density $\bar{\Phi}_{\text{S}}$ we calculate the pair correlation function $g(r)$ by minimizing the density functional $\Omega = \mathcal{F}_{\text{id}} + \int\upd \bvec{r} [\bar{\Phi}_{\text{HS}}+\bar{\Phi}_{\text{S}}+\rho(\bvec{r})(V_{\text{ext}}(\bvec{r})-\mu)]$, where $\mathcal{F}_{\text{id}}$ is the functional for the ideal gas, $\mu$ is the chemical potential and $V_{\text{ext}}$ is the external potential, chosen so that it represents a particle of the SHS fluid at position $r=0$. The resulting density profile $\rho(r)$ gives $g(r) = \rho(r)/\rho_{\text{b}}$, where $\rho_{\text{b}}$ is the bulk fluid density. In Fig.~\ref{fig_gr} we show the case of strong adhesion, $\tau = 0.13$, and compare the FMT results with MC simulations obtained using the algorithm described in \cite{MiFr04}. The regularized $\bar{\Phi}_{\text{S}}$ is superior to the PY-based $\Phi_{\text{S}}$, which is remarkable because $\bar{\Phi}_{\text{S}}$ has been optimized for the 0D limit and not at all for $g(r)$. Other, larger $\tau$, MC simulations for $g(r)$ (see \cite{MiFr04b}) are equally well described by the two FMTs, which give very similar results for $\tau>0.2$. 

\begin{figure}[tbp]
    
  \includegraphics[height = 4.5cm]{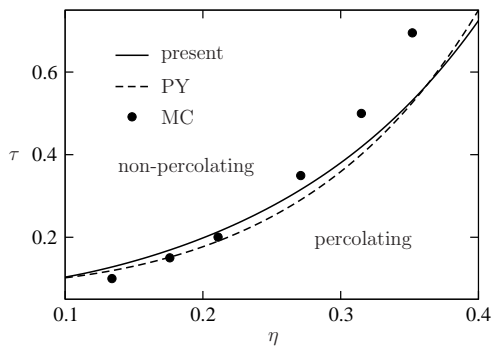} 

  \caption{The percolation line as obtained from MC simulations \cite{Le01} compared to the PY result (dashed line) and the results based on the non-diverging FMT constructed here (solid line).}\label{fig_perc}
  
\end{figure}

Alternatively, an analytical expression for the direct correlation function $c(r)$ can be obtained from the excess free energy $\mathcal{F}_{\text{ex}} = \int \upd\bvec{r}(\bar{\Phi}_{\text{HS}}+\bar{\Phi}_{\text{S}})$ through
\begin{equation}
\label{eq_DCF}
  c(r) = - \frac{\delta^2\mathcal{F}_{\text{ex}}}{\delta \rho(\bvec{r}) \delta \rho(\bvec{r}')} = - \sum_{\alpha, \beta} \frac{\partial^2(\bar{\Phi}_{\text{HS}}+\bar{\Phi}_{\text{S}})}{\partial n_{\alpha} \partial n_{\beta} } \omega_{\alpha \beta}(r) \, ,
\end{equation}
where $\alpha$ and $\beta$ run through $\{ 1, 2, 3, {\bf 1}, {\bf 2}\}$ and $r = |\bvec{r}-\bvec{r}'|$. The partial derivatives of $\bar{\Phi}_{\text{HS}}+\bar{\Phi}_{\text{S}}$ are evaluated in the bulk fluid and derivatives with respect to vectorial quantities $\bvec{n}_1$ and $\bvec{n}_2$ are executed formally while the fact that $\bvec{n}_1=0$ and $\bvec{n}_2=0$ in bulk ensures that the resulting expressions are scalars. The symmetric coefficients $\omega_{\alpha \beta}$ are convolutions of the weight functions that are associated with the respective weighted densities. Therefore $\omega_{\alpha \beta}\equiv 0$ for $r>2R$. For example, $\omega_{33} = \frac{\pi}{12}(4R+r)(2R-r)^2$ is the overlap volume of two spheres with radius $R$ and center-to-center distance $r$. Moreover, $\omega_{23} = \pi R(2R-r)$, $\omega_{13} =-r/8+R/2-R^2/(4r)$, $\omega_{22} =2\pi R^2/r$, $\omega_{12} =R/(4r)+R\delta(2R-r)/8$, $\omega_{\bf 22} = 2\pi R^2/r-\pi r$, and $\omega_{\bf 12} =R/(4r)- R \delta(2R-r)/8$ are required in Eq.~\eqref{eq_DCF}.

In particular, the coefficient of the $\delta$-function in $c(r)$ is $a=-\frac{R}{4}\left(\frac{2\phi_{12}}{1-\eta}+\frac{3\eta\phi_{12}'}{(1-\eta)^2} \right)$. This result can be used to calculate the percolation line which, under the approximations described in \cite{ChGl83}, is $\frac{R}{6a} = \eta$. As long as $\tau$ is not too large, both the classic PY result and the modified FMT describe the percolation threshold as determined in simulations, the latter being somewhat more reliable in the range of intermediate $\tau$ (Fig.~\ref{fig_perc}).

\begin{figure}[tbp]
    
  \includegraphics[height = 4.5cm]{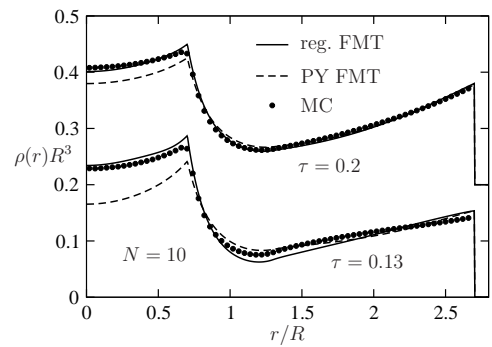} 

  \caption{Density profiles for $N=10$ sticky hard spheres with Baxter parameter $\tau=0.13$ and $\tau=0.2$ in a spherical cavity of radius $R_{\text{cav}}=3.7 R$. The FMT results of the previous PY-based FMT and the present regularized FMT were obtained in the grand canonical ensemble and transformed to the canonical ensemble (with correction of order $1/N^3$) for a comparison with canonical MC simulations that were performed using the algorithm of \cite{MiFr04}.}\label{fig_profiles}
  
\end{figure}

Finally, we consider a confinement scenario for which our theory is expected to be the most relevant. We compare the previous PY-based FMT and the present regularized FMT to canonical MC simulations for a SHS fluid with $\tau=0.13$ and $\tau=0.2$ in a spherical cavity of radius $R_{\text{cav}}=3.7 R$. This system was considered previously \cite{Goetal97} for a hard-sphere fluid ($\tau\to\infty$). To compare the FMTs with the simulations we have to transform the former from the grand canonical ensemble to the canonical ensemble which can be achieved up to corrections of order $1/N^3$ by using the scheme of \cite{Goetal97} that is based on results from \cite{Saetal96}. Fig.~\ref{fig_profiles} shows that the regularized FMT is clearly superior to the previous PY-based FMT, especially in the center of the spherical cavity. Interestingly, while the previous FMT underestimates the density in the center when $\tau$ is small, it overestimates the density in the center when applied to hard spheres; large $\tau$ (see \cite{Goetal97}). The crossover occurs around $\tau=0.5$ where the regularized and the PY-based FMT yield very similar results. In the limit of large cavity radius $R_{\text{cav}}$, planar wall density profiles are recovered; comparison (not shown) with existing MC simulations \cite{JaBr94} shows that the PY-based and the regularized FMT describe the density profiles equally well. In particular, the theories yield identical contact values $\rho_c$ for the density at a hard wall. This results from $\bar{\Phi}_{\text{HS}}+\bar{\Phi}_{\text{S}}$ being equal to  $\Phi_{\text{HS}}+\Phi_{\text{S}}$ in bulk and hence the underlying pressure being the PY compressibility result  $p_{\text{PY}}$ for both theories. Given that both FMTs obey the contact theorem $\rho_c=\beta p_{\text{PY}}$ the contact values of the density must be identical. This illustrates the importance of preserving the bulk fluid properties while removing divergences for highly peaked density distributions in order to obtain a density functional theory that is robust and accurate for a broad spectrum of settings.

\end{document}